\documentclass[pdftex,12pt]{article}
\usepackage{jheppub,amsmath,amsfonts,amsbsy,latexsym,jheppub,amssymb,amscd,amstext}
\usepackage{pst-node}
\usepackage{dsfont}
\usepackage{cool}
\usepackage{overpic,youngtab}
\pdfoutput=1
\usepackage{rotating}
\usepackage{tikz}
\usepackage{slashed}
\UseRawInputEncoding

\def\bpm{\begin{pmatrix}}
\def\epm{\end{pmatrix}}
\def\be{\begin{equation}}

\def\ee{\end{equation}}
\def\bea{\begin{eqnarray}}
\def\eea{\end{eqnarray}}
\def\pd{\partial}

\def\d{\delta}
\def\m{\mu}

\def\n{\nu}
\def\t{\tau}

\def\r{\rho}

\def\xp{x^\prime}

\def\bp{\bar{\phi}}
\def\s{\sigma}
\def\e{\epsilon}

\def\bma{\begin{pmatrix}}
\def\ema{\end{pmatrix}}

\def\bp{\bar{\phi}}

\def\bi{\begin{itemize}}
\def\ei{\end{itemize}}

\def\bp{\bar{\phi}}

\begin{document}

		\vspace*{-1cm}
		\phantom{hep-ph/***} 
		{\flushleft
			{{FTUAM-21-}}
			\hfill{{ IFT-UAM/CSIC-21-16}}}
		\vskip 1.5cm
		\begin{center}
		{\LARGE\bfseries Effective action in elliptic and hyperbolic  spacetimes.}\\[3mm]
			\vskip .3cm
		
		\end{center}

		\vskip 0.5  cm
		\begin{center}
			{\large Enrique \'Alvarez and Jes\'us Anero.}
			\\
			\vskip .7cm
			{
				Departamento de F\'isica Te\'orica and Instituto de F\'{\i}sica Te\'orica, 
				IFT-UAM/CSIC,\\
				Universidad Aut\'onoma de Madrid, Cantoblanco, 28049, Madrid, Spain\\
				\vskip .1cm

				\vskip .5cm
				\begin{minipage}[l]{.9\textwidth}
					\begin{center} 
						\textit{E-mail:} 
					\tt{enrique.alvarez@uam.es},
					\tt{jesusanero@gmail.com}
					\end{center}
				\end{minipage}
			}
		\end{center}
	\thispagestyle{empty}
	
\begin{abstract}
	\noindentŒ 
\end{abstract}

\newpage
\tableofcontents
	\thispagestyle{empty}
\flushbottom

\newpage
 \section{Introduction}
There are not many examples of exact effective actions, even to one loop order, and even for scalar fields. The usual approach (cf. \cite{Alvarez2020} and references therein) only determines the  ultraviolet divergent small proper-time DeWitt coefficients. This leaves undetermined the infrared behavior.

There is however a  theorem \cite{Xu}\cite{Mckean} asserting that whenever the spacetime manifold is such that its Ricci curvature is non-negative, $R_{\m\n}\geq 0$, and the manifold has got maximal volume growth, then the heat kernel corresponding to the ordinary laplacian obeys
\be
\lim_{\t\rightarrow\infty} V\left(\sqrt{\t}\right)\,K\left(x,\xp;\t\right)={\Omega(n)\over (4\pi)^{n/2}}
\ee
where $\Omega(n)$ is the volume of the unit ball in $\mathbb{R}^n$ and  $V\left(\sqrt{\t}\right)$ is the volume of the geodesic ball centered at $\xp$ and radius $\sqrt{\t}$.
The {\em asymptotic volume ratio} is defined as
\be
\lim_{r\rightarrow\infty}{V(r)\over r^n}= \Theta >0
\ee
The fact that $\Theta >0$ is what qualifies for the assertion that the manifold has {\em maximal volume growth}. In fact this is a generalization of a previous theorem by Li and Yau \cite{Li} asserting that with the same hypothesis there should exist a constant $C(\e)$ such that
\be
{1\over C(\e)V\left(\sqrt{\t}\right)}\,e^{-{\s(x,\xp)\over (2-\e/2)\t}}\leq K(x,\xp;\t)\leq {C(\e)\over V\left(\sqrt{\t}\right)}\,e^{-{\s(x,\xp)\over (2+ \e/2)\t}}
\ee

 The situation improves in spacetimes with special amounts of symmetry, where we can find exact expressions for the heat kernel corresponding to the ordinary laplacian\footnote{
  Related computations have been done for the Dirac operator by Camporesi \cite{Camporesi}. See also \cite{Bytsenko}}.
\par
In this work we shall precisely  be concerned with  maximally symmetric spacetimes and their euclidean counterparts. The Riemann tensor obeys
\be
R_{\m\n\r\s}={R\over n(n-1)}\left(g_{\m\r} g_{\n\s}-g_{\m\s}g_{\n\r}\right)
\ee
and the curvature can be positive or negative
\be
R=\pm {n(n-1)\over L^2}
\ee
{\em Elliptic  spacetimes} have got positive curvature. In our conventions, anti-de Sitter spacetime ($AdS_n$) is one such. In Poincar\'e coordinates
\be
ds^2_{AdS_n}={\sum^{n-1}\eta_{ij}dy^i dy^j-L^2dz^2\over z^2}
\ee 
(where as usual, $\eta_{ij} \equiv diag(1, -1
,\ldots,-1
)$). Its euclidean version is the sphere $S_n$, which does not admit Poincar\'e coordinates \cite{AV}, although, as all other spacetimes considered here, does admit stereographic coordinates.
\par
{\em Hyperbolic spacetimes} have got negative curvature. De Sitter ($dS_n$) spacetime falls in this category. In Poincar\'e coordinates
\be
ds^2_{dS_n}={-\sum^{n-1}\d_{ij}dy^i dy^j+L^2dz^2\over z^2}
\ee
with $z$ is a timelike coordinate, the euclidean version reads
\be
ds^2_{EdS_n}={\sum^{n-1}\d_{ij}dy^i dy^j+L^2dz^2\over z^2}
\ee
Correlators, including the energy-momentum tensor in this family of spaces have been thoroughly analyzed in \cite{Osborn} under the hypothesis that those only depend on the invariant arc length, $s$ and its derivatives. Physically this is equivalent to the assumption that the relevant vacuum enjoys all spacetime isometries. 
\par
For timelike geodesics the arc length coincides with the physical {\em proper time}. We shall refrain from using this notation though because we shall use it in Schwinger's sense later.
\par
 We shall also assume that hypothesis (that is, that the the only dependence on coordinates is through the geodesic length) in the present work; this amounts to  demand invariance (or proper behavior) under all conformal isometries \cite{ASG}.
 \par
  When working with lorenztian signature the square of the arc length is not positive semidefinite; it can become zero or even negative. This is in fact the reason why J.L. Synge introduced the {\em world function}, \cite{Synge}, which is essentially the square of the invariant arc length. 
  \par
  Our formulas however remain formally valid with appropiate analytic continuation.
 \section{Hyperbolic Heat kernel}
Acting on functions of the geodesic arc length \cite{Osborn} the laplacian in $\mathbb{H}_n$ reads
\be
\Box={\pd^2\over \pd s^2}+{n-1\over L\tanh\,{s\over L}}{\pd\over \pd s}
\ee
The corresponding  heat equation reads
\be \frac{\partial}{\partial \t}K_{p+1}(\t,s)=\mathcal{D}_p K_{p+1}(\t,s)\ee
where
\be\mathcal{D}_p K_{p+1}(\t,s)=\left(\frac{\partial^2}{\partial s^2}+\frac{p}{L \tanh {s\over L}}\frac{\partial}{\partial s}\right)K_{p+1}(\t,s)\ee
In the following we shall often work with in terms of a dimensionless proper time and dimensionless arc length, ${s\over L}$.

\bi
\item
In the flat limit $s\rightarrow 0$ this reduces to
\be\frac{\partial^2}{\partial s^2}K(\t,s)+\frac{p}{s}\frac{\partial}{\partial s}K(\t,s)=\frac{\partial}{\partial \t}K(\t,s)\ee
and the canonical solution is
\be K(\t,s)=(4\pi\t)^{-(p+1)/2}\exp\left(-\frac{s^2}{4\t}\right)\ee 
this obeys the correct boundary condition
\be
\lim_{\t\rightarrow\infty} K_0(\t,s)=\d^n(x-\xp)\neq \d(s)
\ee
\item
In the opposite limit, $s\rightarrow\infty$ the heat equation reduces to
\be\frac{\partial^2}{\partial s^2}K(\t,s)+p\frac{\partial}{\partial s}K(\t,s)=\frac{\partial}{\partial \t}K(\t,s)\ee
whose general solution is a wave packet composed out of
\be K(\t,s)=\sqrt{\frac{\pi}{\t}}\exp\left(-\frac{s^2}{4\t}-\frac{p s}{2}-\frac{p^2\t}{4}\right)\ee
this does not satisfy the boundary condition at $\t=0$, but this is presumably natural because our approximation is valid for large values of $s$ only.

\item
Let us find a recurrence relation in flat spacetime. This recurrence relation is known in the mathematical literature \cite{Davies}, but our proof stems directly from the heat equation.
Apply the lineal approximation
\be \mathcal{D}^L_p=\frac{\partial^2}{\partial s^2}+\frac{p}{s}\frac{\partial}{\partial s}\ee
with $p$ completely arbitrary 
\be\mathcal{D}^L_p\left( \frac{K_{p-1}'}{s}\right)=\frac{2-p}{s^3}K_{p-1}'+\frac{p-2}{s^2}K_{p-1}''+\frac{1}{s}K_{p-1}'''\ee
where $K'=\frac{\partial K}{\partial s}$, on other hand, let us assume that the function $K$ obeys the heat kernel equation on  $p-2$ dimension, and  derive  one more time
\be \left(\frac{\partial^3}{\partial s^3}+\frac{p-2}{s}\frac{\partial^2}{\partial s^2}-\frac{p-2}{s^2}\frac{\partial}{\partial s}\right)K_{p-1}=\frac{\partial}{\partial \t}K_{p-1}'\ee
obtain
\bea\mathcal{D}^L_p\left( \frac{K_{p-1}'}{s}\right)&&=\frac{1}{s}\frac{\partial}{\partial \t}K_{p-1}'\eea
in conclusion, this implies
\be K_{p+2}(\t,s)=-\frac{1}{2\pi s}\frac{\partial}{\partial s}K_{p}(\t,s)\ee

\par
Of course, in flat spacetime, where we know the full dependence of the heat kernel with the spacetime dimension
\be K_n(s)=\frac{1}{(4\pi\t)^{n/2}}e^{-s^2/4\t}\ee  
there is another trivial recurrence relation
\be
K_{n+1}(s)=-{1\over  s}\sqrt{\t\over \pi}\,{\pd K_n(s)\over \pd s}
\ee

\item
Let us try to generalize this to the hyperbolic case.
Consider the expression (which is independent of the dimension $p$)
\be\mathcal{D}_p\left( \frac{K_{p-1}'}{\sinh s}\right)=\frac{1-p}{\sinh s}K_{p-1}'+\frac{2-p}{\sinh^3 s}K_{p-1}'+\frac{(p-2)\cosh s}{\sinh^2 s}K_{p-1}''+\frac{1}{\sinh s}K_{p-1}'''\ee
if we derive again the heat equation for $p-2$ dimension
\be \left(\frac{\partial^3}{\partial s^3}+\frac{p-2}{\tanh s}\frac{\partial^2}{\partial s^2}-\frac{p-2}{\sinh^2 s}\frac{\partial}{\partial s}\right)K_{p-1}=\frac{\partial}{\partial \t}K_{p-1}'\ee
obtain
\bea\mathcal{D}_p\left( \frac{K_{p-1}'}{\sinh s}\right)&&=\frac{1-p}{\sinh s}K_{p-1}'+\frac{1}{\sinh s}\frac{\partial}{\partial \t}K_{p-1}'=e^{-(1-p)\t}\frac{\partial}{\partial \t}\left(e^{(1-p)\t}\frac{K_{p-1}'}{\sinh s}\right)\nonumber\\\eea
this implies
\be \frac{K_{p-1}'}{\sinh s}=e^{-(1-p)\t}K_{p+1}\ee
because with the heat equation
\bea\mathcal{D}_p\left( e^{-(1-p)\t}K_{p+1}\right)&&=e^{-(1-p)\t}\frac{\partial}{\partial \t}K_{p+1}\eea
finally we have the recurrence relationship
\be K_{p+2}(\t,s)=-\frac{e^{-p\t}}{2\pi\sinh s}\frac{\partial}{\partial s}K_{p}(\t,s)\ee

It is to be stressed that this relationship is a consequence of the heat equation  exclusively. {\em Independently of any  boundary conditions}.
\ei

\subsection{Dimensional reduction from odd to even dimensions.}

The starting point in the recurrence relationship is the formal one-dimensional case.
It can be easily checked that
\be
K_1(\t,s)={e^{-{s^2\over 4 \t}}\over (4\pi\t)^{1/2}}
\ee
it obeys
\be
{\pd^2 \over \pd s^2}K_1(\t,s)={\pd \over \pd \t} K_1(\t,s)
\ee
the engineering dimensions of the heat kernel are determined by its behavior when $\t=0$.
From here on we can determine via the recurrence all odd dimension heat kernels.

The recurrence relationship easily leads to
\be\label{r} K_{2p+1}(\t,s)=\frac{e^{-p^2\t}}{(2\pi)^p}\left(-\frac{1}{\sinh s}\frac{\partial}{\partial s}\right)^p K_1(\t,s)\ee
\par
Here we can see  why dimensional reduction from $n=2p+1$ towards $n=2p$ (what Jacques Hadamard \cite{Hadamard} dubs "the method of descent") does not naively work in this case. 
\par
The reason is clearly that 
\be
p\equiv {n-1\over 2}
\ee
is a fractional number for even $n$, so that we have to take a fractional power of the operator 
\be
{\cal D}\equiv -{1\over \sinh\,s}{\pd\over \pd\,s}
\ee
which  has been worked out for example in \cite{Anker} in the framework of Scr\"odinger's equation, with the result for hyperbolic spaces,
\be
\left(-{1\over \sinh\, s}{\pd\over \pd s}\right)^{n-1\over 2}\,f(s)={1\over \sqrt{\pi}}\int_s^\infty\,dx\,\left(-{1\over \sinh\,x}{\pd\over \pd x}\right)^{n/2} \,f(x)\,{\sinh\,x\over \sqrt{\cosh\,x-\cosh\,s}}
\ee

Let us examine in detail a couple of examples.

\section{Four dimensional hyperbolic space ${\cal H}_4$.}
Our recurrence relation leads to
\bea K_{4}(\t,s)
&&=\frac{\sqrt{2}e^{-{9\t\over 4}-M^2 L^2 \t}}{(4\pi\t)^{5/2}}\int_s^\infty dx\frac{x^2-2\t+2x\t\coth x}{\sinh^2 x}\frac{\sinh x}{\left(\cosh x-\cosh s\right)^{1/2}}e^{-\frac{x^2}{4\t}}\nonumber\\\eea

The effective action (related work is found in e.g. \cite{Schomblond}) is 
\bea V_{eff}[\bar{\phi}]\equiv\int_0^\infty \frac{d\t}{\t} K(\t,s)\eea
for a ${g\over 4!} \phi^4$ self-interaction is then given in the effective potential approximation by
\bea V_{eff}[\bar{\phi}]&&=\frac{\sqrt{2}}{(4\pi)^{2}}\int_s^\infty dx\frac{1}{x^3\sinh^2 x}\frac{\sinh x}{\left(\cosh x-\cosh s\right)^{1/2}}\Big[8+4x\sqrt{9+4M^2L^2}+\nonumber\\
&&+x^2(9+4M^2L^2)+2x\Big(2+x\sqrt{9+4M^2L^2}\Big)\coth x\Big]e^{-x\sqrt{9+4M^2L^2}/2}+{g\over 4!} \bp^4\nonumber\\\eea
where
\be
M^2\equiv m^2+\frac{g}{2}\bp^2
\ee

The minimum is still at $\bp=0$ as long as $m^2\geq 0$.

\section{Five dimensional hyperbolic space ${\cal H}_5$}
Following with recurrence again leads to the five-dimensional heat-kernel
\bea K_{5}(\t,s)s
&&=\frac{e^{-4\t-M^2L^2\t-\frac{s^2}{4\t}}}{(4\pi\t)^{5/2}}\frac{s^2-2\t+2s\t\coth s}{\sinh^2 s}\eea
there is no polynomial interaction of the type $ g \phi^n $ that enjoys dimensionless coupling constant in five dimensions, since the field itself has $[\phi]=3/2$. 

\par
The only potential (besides the mass term) with a positive dimension coupling constant corresponds to a $g\, \phi^3$ interaction with $[g]=1/2$, which we will assume henceforth.
\par
The effective potential, in turn, reads
\bea 
V_{eff}[\bar{\phi}]&&=\frac{1}{(2\pi)^{2}} \frac{e^{-s\sqrt{M^2L^2+4}}}{s^3\sinh^2 s}\Big[2+2s\sqrt{M^2L^2+4}+s^2(M^2L^2+4)+s\Big(1+s\sqrt{M^2L^2+4}\Big)\coth s\Big]+\frac{g}{3!}\bp^3\nonumber\\\eea
where the effective mass in order to compute the effective potential is 
\be
M^2\rightarrow m^2+ g \bp
\ee
and we are making the approximation that $\bp$ is constant
\be
\pd_\m \bp=0
\ee
The minimum is still at the origin $\bp=0$ as long as $m^2\geq 0$.
\section{Elliptic spacetimes.}
Acting on functions of the geodesic arc length \cite{Osborn} the laplacian in $\mathbb{E}_n$ reads
\be
\Box={\pd^2\over \pd s^2}+{n-1\over L\tan\,{s\over L}}{\pd\over \pd s}
\ee
There is a quite similar recurrence in the case of positive curvature, with trigonometric functions taking the place of hyperbolic ones. We shall be brief here. Again we start with the operator of heat kernel over $K$
\be
\mathcal{D}_p\left( \frac{K_{p-1}'}{\sin s}\right)=\frac{p-1}{\sin s}K_{p-1}'+\frac{2-p}{\sin^3 s}K_{p-1}'+\frac{(p-2)\cos s}{\sin^2 s}K_{p-1}''+\frac{1}{\sin s}K_{p-1}'''
\ee
on other hand if derive again the heat equation for $p-2$ dimension
\be \left(\frac{\partial^3}{\partial s^3}+\frac{p-2}{\tan s}\frac{\partial^2}{\partial s^2}-\frac{p-2}{\sin^2 s}\frac{\partial}{\partial s}\right)K_{p-1}=\frac{\partial}{\partial \t}K_{p-1}'\ee
obtain
\bea\mathcal{D}_p\left( \frac{K_{p-1}'}{\sin s}\right)&&=\frac{p-1}{\sin s}K_{p-1}'+\frac{1}{\sin s}\frac{\partial}{\partial \t}K_{p-1}'=e^{-(p-1)\t}\frac{\partial}{\partial \t}\left(e^{(p-1)\t}\frac{K_{p-1}'}{\sin s}\right)\eea
this implies
\be \frac{K_{p-1}'}{\sin s}=e^{-(p-1)\t}K_{p+1}\ee
because with the heat equation
\bea\mathcal{D}_p\left( e^{-(p-1)\t}K_{p+1}\right)&&=e^{-(p-1)\t}\frac{\partial}{\partial \t}K_{p+1}\eea

Finally we get the recurrence relationship
\be \label{re}K_{p+2}(\t,s)=-\frac{e^{p\t}}{2\pi\sin s}\frac{\partial}{\partial s}K_{p}(\t,s)
\ee
Let us work out an example in some detail.
\section{Elliptic five  dimensional space, ${\cal E}_5$.}
The starting point, as well as in the hyperbolic case, is the $n=1$ heat kernel which is common to both cases
\be
K_1={e^{-{s^2\over 4 \t}}\over (4\pi \t)^{1/2}}
\ee
we use the recurrence \eqref{re} with the normal term in mass
\bea K_{5}(\t,s)
&&=\frac{e^{4\t-M^2L^2\t-\frac{s^2}{4\t}}}{(4\pi\t)^{5/2}}\frac{s^2-2\t+2s\t\cot s}{\sin^2 s}\eea
\par
The effective action is defined by
\bea 
V_{eff}[\bar{\phi}]&&=\frac{1}{(2\pi)^{2}} \frac{e^{-s\sqrt{M^2L^2-4}}}{s^3\sin^2 s}\Big[2+2s\sqrt{M^2L^2-4}+s^2(M^2L^2-4)+s\Big(1+s\sqrt{M^2L^2-4}\Big)\cot s\Big]+\frac{g}{3!}\bp^3\nonumber\\\eea
where the effective mass in order to compute the effective potential is 
\be
M^2\rightarrow m^2+ g \bp
\ee
in the conclusions we explain in detaill this result, for the effective action on elliptic spacetime.

  \section{Conclusions}
 
 It is remarkable that the effective action for massless scalars in elliptic spacetimes (which are compact when euclidean) is IR divergent, whereas in (non compact even when euclidean)  hyperbolic spacetimes it is not. 
 \par
 Actually, there is always a minimal value for the effective mass above  which the effective action does converge.
 For example, for five dimensional elliptic spaces, IR divergences in the effective action disappear whenever
 \be
 M^2\equiv m^2+g\bp \geq {4\over L^2}
 \ee
 The reason for this divergence in the massless case might be related to the presence of antipodal points. The reason is that the laplacian in elliptic spaces
 \be
\Box={\pd^2\over \pd s^2}+{n-1\over L\tan\,{s\over L}}{\pd\over \pd s}
\ee
is singular not only when $s=0$, but also whenever
\be
s=n \pi L
\ee
for any admissible integer $n$. Usually $n=1$ corresponds to the geodesic distance to the antipodal point.
\par
 This fails to happen in hyperbolic spaces, where the hyperbolic tangent only vanishes when $s=0$. 
 \par
 This is the reason why Schr\"odinger \cite{Schrodinger} proposed already in 1957 the elliptic interpretation, where all antipodal points are identified. 
 \par
 While suggestive, the fact that the spacetime $dS_n/\mathbb{Z}_2$ is not orientable poses many physical problems.
 \section{Acknowledgements}
One of us (EA) is grateful for useful correspondence with Jos\'e Gracia-Bond\'{\i}a.  This work has been supported in part by AEI grant PID2019-108892RB-I00/AEI/10.13039/501100011033 as well as from the Spanish Research Agency (Agencia Estatal de Investigacion) through the grant IFT Centro de Excelencia Severo Ochoa SEV-2016-0597, and the European Union's Horizon 2020 research and innovation programme under the Marie Sklodowska-Curie grants agreement No 674896 and No 690575. We have also been partially supported by FPA2016-78645-P(Spain). This project has received funding /support from the European UnionÕs Horizon 2020 research and innovation programme under the Marie Sklodowska -Curie grant agreement No 860881-HIDDeN.
\newpage
\appendix

\newpage


\begin{thebibliography}{99}
\bibitem{Alvarez2020}
E.~Alvarez,
``Windows on Quantum Gravity,''
[arXiv:2005.09466 [hep-th]].
Fortschritte der Physik
2020-10-06 | journal-article
DOI: 10.1002/prop.202000080
\bibitem{Bytsenko}
  A.~A.~Bytsenko, S.~D.~Odintsov and S.~Zerbini,
  ``The Effective action in gauged supergravity on hyperbolic background and induced cosmological constant,''
  Phys.\ Lett.\ B {\bf 336} (1994) 355
  doi:10.1016/0370-2693(94)90545-2\\
  A.~A.~Bytsenko, S.~D.~Odintsov and S.~Zerbini,
  ``The Large distance limit of the gravitational effective action in hyperbolic backgrounds,''
  Class.\ Quant.\ Grav.\  {\bf 12} (1995) 1
   Erratum: [Class.\ Quant.\ Grav.\  {\bf 12} (1995) 2355]
  doi:10.1088/0264-9381/12/1/002
  [hep-th/9410112].\\
  A.~Bytsenko, K.~Kirsten and S.~Odintsov,
  ``Selfinteracting scalar fields on space-time with compact hyperbolic spatial part,''
  Mod.\ Phys.\ Lett.\ A {\bf 8} (1993) 2011
  doi:10.1142/S0217732393001720
  [hep-th/9303061].

\bibitem{Xu}
Guoyi Xu.
"Large time behavior of the heat kernel"
arXiv:1310.2382v1 [math.DG]\\
\bibitem{Mckean}
H. P. McKean,
"An upper bound to the spectrum of $\Delta$ on a manifold of negative curvature,
J . Diff Geom. 4 (1970) 359-366


\bibitem{Li}
Peter Li and Shing Tung Yau,
"On the parabolic kernel of the Schr\"odinger operator ",
Acta Math, 156,(1983) 153.
\bibitem{Camporesi}
R.~Camporesi,
``The Spinor heat kernel in maximally symmetric spaces,''
Commun. Math. Phys. \textbf{148} (1992), 283-308
doi:10.1007/BF02100862\\
R.~Camporesi and A.~Higuchi,
``On the Eigen functions of the Dirac operator on spheres and real hyperbolic spaces,''
J. Geom. Phys. \textbf{20} (1996), 1-18
doi:10.1016/0393-0440(95)00042-9
[arXiv:gr-qc/9505009 [gr-qc]].
\bibitem{AV}
E.~Alvarez and R.~Vidal,
``Eternity and the cosmological constant,''
JHEP \textbf{10} (2009), 045
doi:10.1088/1126-6708/2009/10/045
[arXiv:0907.2375 [hep-th]].
\bibitem{Osborn}
B.~Allen and C.~A.~Lutken,
``Spinor Two Point Functions in Maximally Symmetric Spaces,''
Commun. Math. Phys. \textbf{106}, 201 (1986)
doi:10.1007/BF01454972\\
B.~Allen and T.~Jacobson,
``Vector Two Point Functions in Maximally Symmetric Spaces,''
Commun. Math. Phys. \textbf{103}, 669 (1986)
doi:10.1007/BF01211169\\
H.~Osborn and G.~M.~Shore,
``Correlation functions of the energy momentum tensor on spaces of constant curvature,''
Nucl. Phys. B \textbf{571} (2000), 287-357
doi:10.1016/S0550-3213(99)00775-0
[arXiv:hep-th/9909043 [hep-th]].
\bibitem{ASG}
E.~Alvarez and R.~Santos-Garcia,
``CFT in Conformally Flat Spacetimes,''
Phys. Rev. D \textbf{101}, no.12, 125009 (2020)
doi:10.1103/PhysRevD.101.125009
[arXiv:2001.07957 [hep-th]].\\
	 \bibitem{Synge}
Synge, J.L.,
Relativity: The General Theory,
(North-Holland, Amsterdam, 1960).
\bibitem{Davies}
E.B. Davies,
"Heat kernels and spectral theory",
(cambridge University press, 2007)
\bibitem{Hadamard}
Jacques Hadamard,
"Lectures on Cauchy's problem in linear partial differential equations",
(Dover)

\bibitem{Anker}
Jean-Philippe Anker, Vittoria Pierfelice ,
"Nonlinear Schršdinger equation on real hyperbolic spaces",
Ann. I. H. PoincarŽ Ð AN 26 (2009) 1853Ð1869

\bibitem{Schomblond}
C.~Schomblond and P.~Spindel,
``Unicity Conditions of the Scalar Field Propagator Delta(1) (x,y) in de Sitter Universe,''
Ann. Inst. H. Poincare Phys. Theor. \textbf{25} (1976), 67-78\\
J.~S.~Dowker and R.~Critchley,
Phys. Rev. D \textbf{13} (1976), 224
doi:10.1103/PhysRevD.13.224
T.~Inami and H.~Ooguri,
``One Loop Effective Potential in Anti-de Sitter Space,''
Prog. Theor. Phys. \textbf{73} (1985), 1051
doi:10.1143/PTP.73.1051\\
C.~P.~Burgess and C.~A.~Lutken,
Phys. Lett. B \textbf{153} (1985), 137-141
doi:10.1016/0370-2693(85)91415-7\\


\bibitem{Schrodinger}
E.~Schrodinger,
``Expanding universes,''\\
(Cambridge, 1956)
M.~K.~Parikh, I.~Savonije and E.~P.~Verlinde,
``Elliptic de Sitter space: dS/Z(2),''
Phys. Rev. D \textbf{67} (2003), 064005
doi:10.1103/PhysRevD.67.064005
[arXiv:hep-th/0209120 [hep-th]].

\end{thebibliography}
\end{document}